\documentclass[twocolumn,amssymb,footinbib, nobibnotes, aps, pra, reprint, a4paper]{revtex4-2}

\usepackage[utf8]{inputenc}
\usepackage{alphabeta}
\usepackage[LGR,T1]{fontenc}
\usepackage[british]{babel}
\usepackage{alphabeta}

\usepackage{mathtools}

\usepackage{physics}

\usepackage{xspace}

\usepackage{hyperref}

\usepackage[capitalise]{cleveref}

\makeatletter
\AddToHook{cmd/appendix/before}{\def\cref@section@alias{appendix}\def\cref@subsection@alias{appendix}}
\makeatother

\usepackage{nicefrac}

\usepackage[expansion=false, protrusion = true]{microtype}

\usepackage[autostyle=true]{csquotes}

\ifLuaTeX
    \usepackage{unicode-math}
    \newcommand{\bm}[1]{\symbf{#1}}

\else
    \usepackage{amssymb}

    \usepackage{bm}
\fi

\usepackage{xcolor}

\newcommand{\mv}[1]{\mathbf{#1}}

\newcommand{\ad}{\ensuremath{a_{\mathrm{d}}}}
\newcommand{\ea}{\ensuremath{a_{\mathrm{e}}}}

\newcounter{myfigpanel}[figure]
\newcounter{myfigpanelonly}[figure]

\newcommand{\panelletter}[1]{\refstepcounter{myfigpanel}\label{#1}\refstepcounter{myfigpanelonly}\label{onlyletter:#1}\Alph{myfigpanel}}
\newcommand{\panel}[1]{(\protect\panelletter{#1})}

\crefalias{myfigpanel}{figure}
\crefname{myfigpanelonly}{panel}{panels}

\let\origcaption\caption
\let\caption\undefined
\DeclareRobustCommand{\caption}[1]{\origcaption{\protect\setcounter{myfigpanel}{0}\protect\setcounter{myfigpanelonly}{0}#1}}

\usepackage{graphicx}

\newcommand\MPIDS{\affiliation{Max Planck Institute for Dynamics and Self-Organization, Göttingen, Germany}}

\makeatother
\begin{document}

\title{Order and shape dependence of mechanical relaxation in proliferating active matter}
\author{Jonas Isensee}
\MPIDS
\author{Lukas Hupe}
\MPIDS
\author{Philip Bittihn}
\email{philip.bittihn@ds.mpg.de}
\MPIDS

\begin{abstract}
Collective dynamics in proliferating anisotropic particle systems arise from an interplay between growth, division, and mechanical interactions, often mediated by particle shape.
In classical models of prolate, rod-like growth, flow-induced alignment and division geometry reinforce one another, leading to robust nematic order under confinement.
Here we introduce a complementary regime by considering smooth convex particles whose geometry can be oblate for part or all of their growth cycle, creating a tunable competition between these two alignment mechanisms.
Using agent-based simulations of elliptical and rounded-rectangular particles in both channel and open-domain geometries, we systematically vary the division aspect ratio to span regimes of cooperation and competition between ordering cues.
We find that oblate growth can reverse classical flow-alignment, destabilize microdomain formation in intermediate regimes, and open up new regimes with modified microdomain dynamics in free expansion and sustained orientation dynamics in channel geometry.
These findings are explained by an order- and shape-dependent mechanical relaxation interpretation that is supported by explicit measurements. This sheds new light on the available relaxation pathways and therefore provides key ingredients for effective descriptions of collective anisotropic proliferation dynamics.
\end{abstract}

\maketitle
\section{Introduction}

Cell growth and division are central to living systems and proliferating active matter. Conceptual challenges, such as non-conserved particle numbers, leave many aspects of proliferating systems poorly understood~\cite{hallatschekProliferatingActiveMatter2023}.
Nonetheless, numerical simulations of particle-based models have been shown to give insight into specific dynamics and inspire novel approaches.
Here, we are most interested in volume growth, which is often only one of several features of real-world systems leading to complex structures, among others such as reductional division, cell migration, and mechano-chemical signaling.
Understanding these processes requires theoretical and experimental approaches, including self-similar division dynamics~\cite{lishIsovolumetricDividingActive2024}, membrane mechanics~\cite{agudo-canalejoPatternFormationCurvatureinducing2017,zwickerGrowthDivisionActive2017,scarpaActomyosinDrivenTensionCompartmental2018}, mechanical effects of the division process~\cite{campinhoTensionorientedCellDivisions2013,doostmohammadiCelebratingSoftMatters2015}, morphogen gradients~\cite{bryantRelationshipGrowthPattern2016}, and stress measurements~\cite{moritaPhysicalBasisCoordinated2017,vosCharacterizingIntracellularMechanics2024,muenkerAccessingActivityViscoelastic2024}.

One biological inspiration for our theoretical work are rod-shaped bacteria which have been studied in experiments using suitable strains of \emph{Escherichia coli}~\cite{volfsonBiomechanicalOrderingDense2008, boyerBucklingInstabilityOrdered2011,dellarcipreteGrowingBacterialColony2018,youGeometryMechanicsMicrodomains2018} or \emph{Bacillus subtilis}~\cite{fortuneBiofilmGrowthElastic2022,basaranLargescaleOrientationalOrder2022}.
Confined to a quasi two-dimensional monolayer and placed into a channel geometry, the growing bacteria reliably end up oriented along the channel.
When allowed to expand freely on a substrate, these bacteria form an exponentially expanding circular colony exhibiting local alignment structure before the central pressure eventually suffices to kickstart an escape into the third dimension~\cite{berozVerticalizationBacterialBiofilms2018,hartmannEmergenceThreedimensionalOrder2019,storckVariableCellMorphology2014}.
What makes this behavior particularly interesting to analyze is that it can be easily reproduced using numerical simulations of growth and dividing rod-shaped particles as has been shown in literature~\cite{winkleModelingMechanicalInteractions2017,youConfinementinducedSelforganizationGrowing2021,isenseeStressAnisotropyConfined2022,langeslayStressAlignmentResponse2024,ratmanSpontaneousSpatialSorting2025}.
Since simple and purely mechanical models can reproduce key features of the emergent orientation dynamics, they must already capture most of the relevant microscopic physics involved.

In these systems, division geometry and alignment are coupled, often reinforcing each other, allowing steric interactions and anisotropy to drive nematic order, defect dynamics, and shear flows~\cite{marchettiHydrodynamicsSoftActive2013,doostmohammadiActiveNematics2018}.
This can lead to rich dynamics featuring the growth and breakup of microdomains~\cite{youGeometryMechanicsMicrodomains2018,isenseeSensitiveParticleShape2025} and unexplained features like the stress anisotropies reported in Ref.~\cite{isenseeStressAnisotropyConfined2022}.

Seeking to understand these dynamics, models have often focused on prolate particles that grow and divide along their long axis~\cite{grantRoleMechanicalForces2014}, with each cycle reinforcing nematic order under confinement.
In an extension of the work in Ref.~\citenum{isenseeSensitiveParticleShape2025}, which introduced a generalization of the typical rod-shape to study tip-shape variations, we present an oblate variant of the rod-shapes and a separate approach using elliptical particles.
While such shapes are not common in naturally dividing cells, introducing them in a controlled model setting allows us to study a qualitatively new situation in which the two main sources of ordering, namely flow-induced alignment due to particle shapes and alignment through proliferation can act in opposition.
This framework enables us to probe how order emerges, reorganizes, or breaks down when these drivers compete rather than reinforce each other, offering a clean physical setting to study frustration and reorientation in active nematic-like systems.

By connecting particle-scale mechanics to orientational order we establish the order-dependent packing fraction as a valuable coarse-grained property and gain insight on inhibition of viscous relaxation through analysis of local comoving velocity variations.
The resulting physical insights are applicable across a broad class of proliferating anisotropic systems, irrespective of whether such oblate regimes naturally occur, and highlight how even subtle changes in shape can qualitatively alter emergent collective dynamics.

\section{Methods}

We perform agent-based simulations in which each particle is described by its position and orientation in space and its internal life-cycle that determines its instantaneous shape and size.
Particle motion arises solely from steric repulsion with other particles and confining boundaries; no explicit self-propulsion or adhesive forces are imposed.
Repulsive forces appear when growth or motion brings particles into contact such that their respective shape outlines overlap.
In this case, a force acts normal to the contact plane, as is illustrated in \cref{pan:fig1_sketches}.

At each simulation step, the translational and rotational motion of all particles is determined by integrating the overdamped equations of motion
\begin{equation}
 \dot{\mathbf{r}}_i = \mu_t\mathbf{F}_i, \quad
 \dot{\theta}_i =\mu_r \tau_i ,
\end{equation}
where $\mathbf{F}_i$ and $\tau_i$ are the net force and torque on particle $i$, and $\mu_t$, $\mu_r$ are translational and rotational mobility coefficients that depend on the particle size.

Central design criteria for particle shapes in this work are that they should have nematic symmetry and be able to extend along one axis to explore the resulting collective dynamics.
The most common shape used in the literature, that we build upon, is the rod model for nematic bacteria.
These consist of a fixed-width rectangular body with half-circle caps at opposing ends as shown in \cref{pan:fig1_sketches}.
A necessary requirement for all models with (indefinite) growth and binary divisions is that particle volume approximately doubles during its life-time.
In the case of the rod model, the width stays fixed while the length increases to reach its division aspect $\ad$ at which point the particle instantaneously divides in the middle to produce two new particles.
Each of these gets assigned a growth rate $γ$ from the uniform distribution on the interval $[\tfrac{3}{4}, \tfrac45]$.
Based on the calculations presented in Ref.~\cite{isenseeStressAnisotropyConfined2022}, we know that the growth progress $g\in [0,1]$, which tracks the cells' internal life cycle, has a population average of $\langle g\rangle \approx 0.4414$.
Consequently, we can expect an average particle aspect ratio $\ea = \tfrac{1+\langle g\rangle}{2}\ad$ which will become a useful quantity below.

Here, we introduce a generalization of this model to allow us to simulate (oblate) particle aspect ratios $a < 1$ where the non-growing width is larger than the growing length for part or all the growth cycle.
Such particles can be constructed as rounded rectangles using the same construction as described in Ref.~\cite{isenseeSensitiveParticleShape2025}.
They consist of a central rectangle flanked by semicircular (or quarter-circle) caps of separately set curvature radius $R_c$.
This enables construction of shapes ranging from prolate ($\ad \geq 2$) rod-like morphologies to flattened, oblate objects.
To keep the space of possible shapes simple, we limit ourselves to the shapes with the maximal geometrically permissible constant corner rounding $R_c = \min(1,\ad / 2)W/2$ with width $W$, as illustrated in \cref{pan:fig1_sketches}.
The width $W$ itself is defined via the approximate cell area which we require to be $1 = W^2 \ad/2$ at birth.

Providing a contrast to the rod-like shapes, elliptical particles are smoothly rounded at all times.
Similar to before, elliptical particles are parameterized by a non-growing width that defines the first semi-axis and a growing length giving the second semi-axis that need not be larger than the width.
Repulsion forces are normal to the particle outline and scale with the overlap as illustrated in \cref{pan:fig1_sketches} and are defined in more detail in \cref{app:repulsionforces}.

We run simulations in two distinct geometries: straight channels with periodic boundaries along the longitudinal direction, and open circular domains, i.e. \emph{disks}.
In both geometries, simulations begin with a small number of randomly oriented particles placed in the central region.
The system is evolved until a steady state is reached where the statistical properties of order and packing fluctuate about constant mean values.

\begin{figure}
    \centering
    \includegraphics[width=\linewidth]{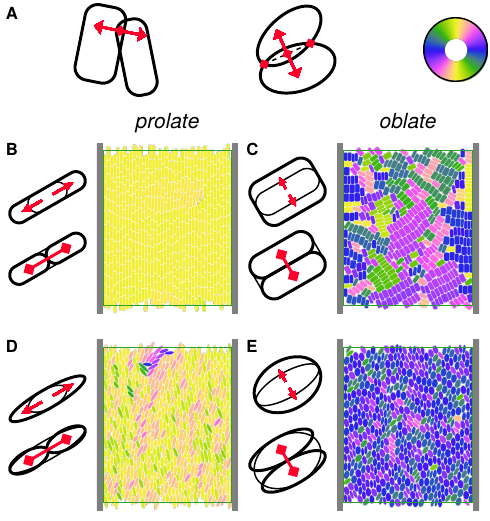}
    \caption{\label{fig:fig1}%
    \panel{pan:fig1_sketches} Sketches of steric repulsion for
    elliptical and rod-shaped particles.
    \panel{pan:fig1_prolate_rod}, \panel{pan:fig1_oblate_rod}, \panel{pan:fig1_prolate_ellipse}, \panel{pan:fig1_oblate_ellipse}
     Snapshots of particle simulations in channel geometry using either prolate or oblate particle shapes as indicated by the accompanying sketches. Particles are colored by their nematic orientation as indicated in the color wheel.
    }
\end{figure}

\section{Results}
\subsection{Channel simulations - imposed shear flow}
The channel geometry provides a controlled setting that imposes shear flow. Flow alignment becomes readily observable as has been shown and studied multiple times for regular ($\ad \geq 2$) rod-shaped particles~\cite{volfsonBiomechanicalOrderingDense2008,boyerBucklingInstabilityOrdered2011,youConfinementinducedSelforganizationGrowing2021,winemanViscoelasticSolids2020}.
This basic case is demonstrated in \cref{pan:fig1_prolate_rod}.
Switching to elliptical particles in \cref{pan:fig1_prolate_ellipse}, we find a very similar picture of channel alignment, albeit lacking the highly ordered columnar structure present in simulations of rod-shaped particles of the same division aspect $\ad$.

Moving to $\ad < 2$, we find that elliptical particles transition from classical alignment (parallel to channel axis) at large $\ad$ to reversed alignment (perpendicular) at small (oblate) $\ad$ seen in \cref{pan:fig1_oblate_ellipse}.
Here, particles are predominantly oriented such that their long axis
points toward the channel exit and their growth points into the confined direction.

This growth is continuously compensated by particle rearrangements easily possible due to the smoothly rounded elliptical shapes.
The same is not true for oblate rods as shown in \cref{pan:fig1_oblate_rod}.
These shapes are dominated by flat sides that inhibit local rearrangements.
Instead, one observes an apparent competition of opposing alignment effects leading to the continuous formation and breakup of microdomains within the channel.

\begin{figure}[htbp]
    \centering
    \includegraphics[width=\linewidth]{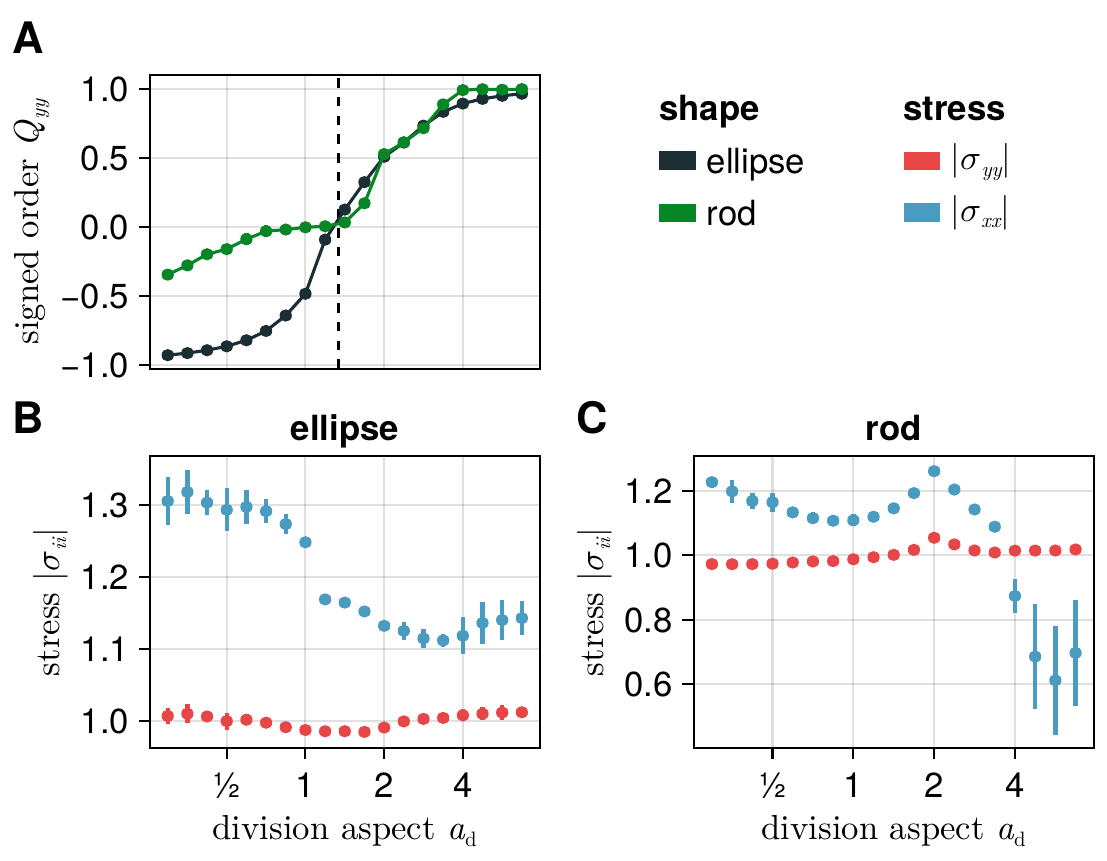}
    \caption{\label{fig:fig2}%
    \panel{pan:fig2_signed_order} Central order parameter $Q_{yy}$ and
    \panel{pan:fig2_ellipsestresschannel}  stresses $|\sigma_{xx}|$ and $|\sigma_{yy}|$ in channel geometry for both elliptical and \panel{pan:fig2_rodstresschannel} rod-shaped particles for varied division aspect $\ad$. Errorbars indicate variation between different initial conditions.
    }
\end{figure}

The steady state averages of the regimes are analyzed quantitatively for varied division aspect $\ad$ in \cref{fig:fig2}.
For quantifying orientational order we use the nematic tensor $Q_{ij} = \langle 2n_{i}n_{j}-\delta_{ij}\rangle$ with $\bm{n}$ the normalized orientation vector of each particle.
In the component $Q_{yy}$ a value of 1 corresponds to maximal alignment along the channel---the growing axis of particles points to the channel outlets---and a value of -1 is maximal order orthogonal to the channel.
This is shown in \cref{pan:fig2_signed_order}.
For prolate rods we see the known and previously characterized regime of partial channel alignment followed by maximal order at large $\ad > 4$~\cite{isenseeStressAnisotropyConfined2022}.
For elliptical particles, the behavior is similar but approaches maximal order more smoothly.

Oblate particles on the other hand generally generate orthogonal alignment $Q_{yy} < 0$.
The transition between the two regimes is found to be close to
$\ad \approx \nicefrac{4}{3}$ shown dashed.
This is the parameter at which particles spend exactly equal amounts of time oblate $a<1$ as prolate $a>1$.
Oblate rods remain largely disordered on average and only generate small amounts of alignment in the channel, while oblate elliptical particles approximately mimic the curve for prolate ellipses at corresponding inverse aspect ratios.
They smoothly approach maximally orthogonal order.

This switch in dynamics between $a>1$ and $a<1$ also manifests in the average stress tensor.
Figs.~\ref{pan:fig2_ellipsestresschannel} and \ref{pan:fig2_rodstresschannel} show the principal stress components $\sigma_{xx}$ and $\sigma_{yy}$ approximated in the channel center for both elliptical and rod-shaped particles.

The vertical stress quantified by $σ_{yy}$ is generated by substrate friction that opposes flow within the channel and toward the outlets.
It depends on the particle growth rates, the particle packing density, which both have well-defined averages, and, trivially, on the mobility coefficients which are kept fixed.
We therefore show all measured stresses normalized by the maximal stress $|σ_{yy}|$ for each model across all values $\ad$.

Keeping this in mind, we turn to the measurements of the stress $|\sigma_{xx}|$.
For elliptical particles in \cref{pan:fig2_ellipsestresschannel} we see that strongly oblate growth causes up to $30\%$ in excess stress $|\sigma_{xx}|$ which first slowly decreases and then abruptly jumps to lower values between $10-15\%$ when the division aspect becomes $\ad > 1$.

The portion of the $|\sigma_{xx}|$ curve for rod-shaped particles in \cref{pan:fig2_rodstresschannel} with $\ad \geq 2$ was already reported and discussed in Ref.~\citenum{isenseeStressAnisotropyConfined2022}.
It features an approximately linear decrease for  $2 < \ad < 4$, followed by a continued downward trend to negative stress anisotropy where $|\sigma_{xx}| < |\sigma_{yy}|$.

At lower division aspect $\ad< 2$ we find a local minimum of $|σ_{xx}|$ near $\ad=1$ followed by an increasingly steep increase toward smaller $\ad$ that is accompanied by the emergence of orthogonal order $Q_{yy} < 0$.

\subsection{Microdomain dynamics and spontaneous shear flow}

\begin{figure*}
    \centering
    \includegraphics[width=1\linewidth]{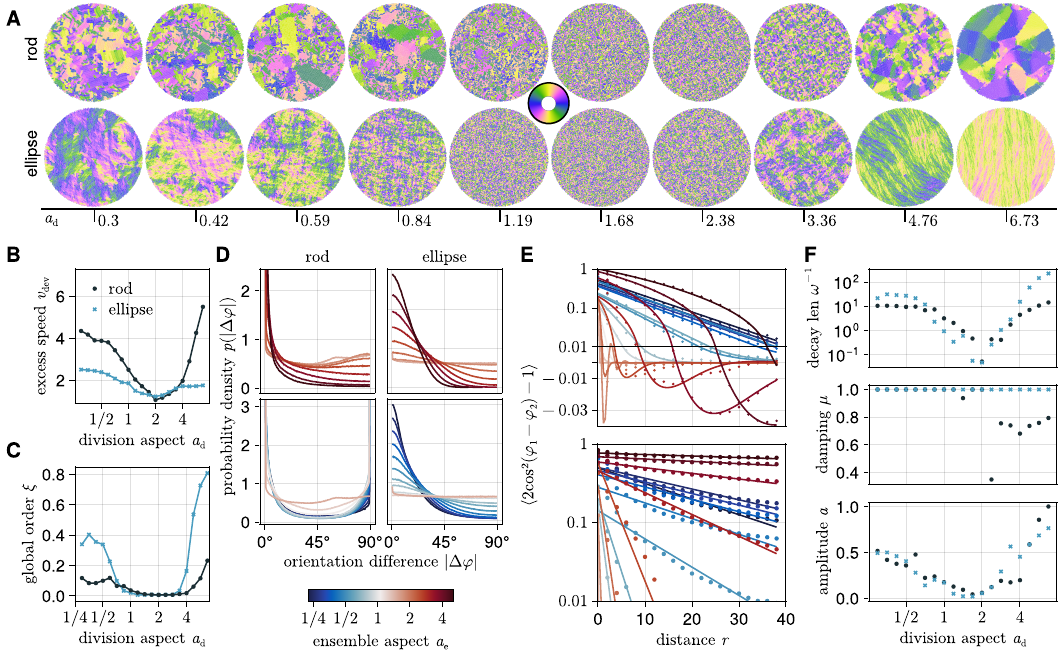}
    \caption{\label{fig:statistics}%
    \panel{pan:many_snapshots} Simulation snapshots for varied particle shapes as labelled.
    \panel{pan:stats_excess_vel} Average instantaneous velocity deviation from the mean-field flow.
    \panel{pan:stats_global_order} Average global instantaneous scalar order parameter $\xi=\sqrt{-\det Q}$.
    \panel{pan:stats_pairwise} Pairwise orientation statistics between touching particles. Data is split into different panels to aid readability.
    \panel{pan:stats_correlation_rod} Two-point orientation correlation for rods (top) and 
    ellipses (bottom) as markers and approximations as a damped oscillation as solid lines.
    \panel{pan:stats_correlation_params} Fit parameters of damped oscillator approximation.
    }
\end{figure*}

Dynamics are generated at the microscopic level by steric repulsion between particles.
This causes motion and rotation depending on the concrete configurations.
In the channel geometry discussed above, the confinement imposed a clear directional bias on the dynamics fully dictating the average expansion flow.
Here, we remove the directional bias by considering a disk-shaped simulation domain with absorbing boundaries at the perimeter.
Given substrate friction and a homogeneously filled domain, one should expect a radially symmetric expansion flow on average.
However, at the microscopic level the symmetry is always broken due to the directed growth of individual particles.
How this symmetry-breaking transfers to many-particle length scales and how the resulting dynamics depend on particles shape can be studied in this setup.
Example snapshots of the dynamics are shown in \cref{pan:many_snapshots}.

Connecting to previous work~\cite{youGeometryMechanicsMicrodomains2018, isenseeSensitiveParticleShape2025}, we find that the coupling between directed growth and local parallel alignment can give rise to microdomains—finite-size regions with maximal local order.
The size distribution of these structures depends sensitively on the particle shape and, here, the largest clusters are observed for rods with the highest $\ad$ tested.
No microdomains can be observed for $\ad<4$ until a new oblate variant of microdomains appears for $\ad < 1$.

Elliptical particles on the other hand do not show stable microdomains regardless of division aspect.
Instead, they exhibit only partial orientational order but with correlations over large distances.
This is qualitatively evident from the snapshots in \cref{pan:many_snapshots} and quantified in more detail in the following.

Due to surface friction and near incompressibility, we expect the average velocity of a disordered aggregate to be the purely radial field
\begin{align*}
    \bm{v}(\bm{x}) = (\partial_rv_r) \bm{x} \label{eq:isotropicexp}
\end{align*}
with the coordinate origin at the center of the domain and average radial velocity gradient
\begin{align*}
    \partial_rv_r = \left\langle \frac{\bm{v}_i\cdot\hat{\bm{e}}_r}{||\bm{x}_i||}\right\rangle \approx \left\langle \frac{(\bm{x}_i^{t+τ}-\bm{x}_i^{t})\cdot\hat{\bm{e}}_r}{τ||\bm{x}_i^{t+τ}+\bm{x}_i^{t}||/2}\right\rangle
\end{align*}
which can be estimated from simulations using a finite difference velocity and averaging over time $t$ and all particles enumerated with index $i$ and instantaneous position $\bm{x}_i^{t}$.
As averaging lag time we use $τ=0.2$ in units of the mean particle lifetime.
Deviations from the mean flow are expressed as
\begin{align}
    v_{\mathrm{dev}} = \left\langle ||\bm{v}_i - (\partial_rv_r) \bm{x}_i||\right\rangle
\end{align}
which we can use to quantify how particle shape and nematic order are reflected in the local flow fields.
Results for various parameters are displayed in \cref{pan:stats_excess_vel}.

For both particle models we find very similar and minimal speed deviations for division aspect values close to $\ad = 2$.
From there it increases in both directions.
However, while there are steep increases for rods that are a signature of the micro-domain dynamics, the changes with elliptical particles are much more gradual and appear to saturate for both oblate and prolate aspect ratios.
In contrast, we see large values for the average of the global instantaneous scalar order parameter for anisotropic elliptical particles in \cref{pan:stats_global_order}, while global order is significantly suppressed for rod-shaped particles.

At the microscopic level we see that elliptical particles do not align with neighboring particles as well as rod-shaped particles.
This is shown in \cref{pan:stats_pairwise} in the form of pairwise alignment statistics gathered across space and time from the simulations.
Roundish particles, those with $1 < \ad < 2$, generate the widest spread in relative orientations and more anisotropic shapes, both oblate and prolate, generate more pronounced pairwise alignment.
However, while elliptical particles only generate a smoothly peaked distribution, rods feature strong peaks near parallel alignment and for $\ad < 2$ also show a second peak at $90\deg$ indicating tip-side alignment.

Despite this lack in strong alignment of ellipses, we find large instantaneous global order and this similarly manifests in the reliable observation of domain-scale orientation correlation with sufficiently anisotropic shapes, meaning either $\ad \gg 1$ or $\ad^{-1} \gg 1$.
A two-point orientation correlation of the form $\left\langle 2\cos^2(\varphi_i - \varphi_j)-1\right\rangle_{r=|\mv{x}_i-\mv{x}_j|},$
where $\varphi_i$ and $\varphi_j$ are particle orientations and the average is computed over particle pairs whose center positions $\mv{x}_i$ are located at euclidean distance $r$ from one another, is shown in \cref{pan:stats_correlation_rod}.
Its fit parameters in the terminology of a damped oscillation
\begin{align}
    c(r) \approx a e^{-\omega \mu r}\cos\left(\omega r\sqrt{1-\mu^2}\right) \label{eq:correlationfitfunc}
\end{align}
with amplitude $a$, spatial frequency $\omega$ and relative damping $\mu$ are displayed in \cref{pan:stats_correlation_params}.
From this we find that configurations of elliptical particles follow an overdamped correlation decay that is realized using critical damping of $μ=1$ and whose decay length $ω^{-1}$ scales strongly with the division aspect $\ad$.

This is an interesting observation and an excellent indication for the origin of the large-scale orientational correlation in growing elliptic particles.
We attribute it to a remarkably weak coupling between individual particle growth and the surrounding shear-like flow.
Orientation patterns magnify in time by growth and persistence of orientation at cell division and as directed growth does not appear to induce meaningful amounts of directed extension at larger scales, these orientation patterns can evolve on much larger time scales than their rod-shaped counterparts.

\subsection{Order-dependent packing fraction}
\begin{figure}
    \centering
    \includegraphics[width=\linewidth]{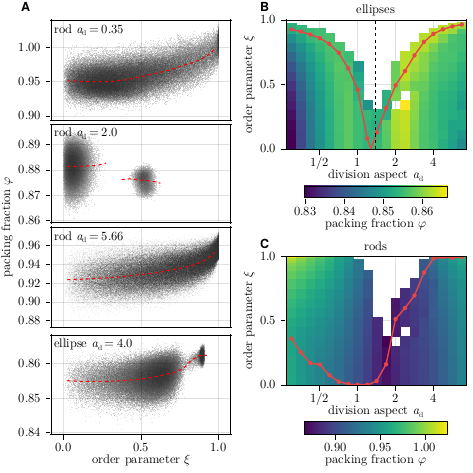}
    \caption{\label{fig:packing}%
    \panel{pan:pointclouds} Packing fraction as a function local order for varied division aspect $\ad$ and particle model as labelled.
    \panel{pan:packing_heatmap_ellipse} Histograms of observed order--packing pairs as heatmap over the division aspect ratio for elliptical particles.
    \panel{pan:packing_heatmap_rod} Analogous to \ref{pan:packing_heatmap_ellipse} for rod-shaped particles.
    }
\end{figure}

Particles under compression move locally into the nearest configurations that use the available space more efficiently thus reducing the pressure.
Well-known examples include monodisperse spheres arranging in a hexagonal close-packed lattice and squares tending to stack.
While crystal lattices are not achievable for the polydisperse mixture of growing particles, we still see orientational order with nematic symmetry.

Here, we show that the dynamics select a subset of the uncountable set of possible particle configurations which we can meaningfully approximate into a single functional relationship between a local nematic order parameter and a packing fraction.
To do so, we divide simulation snapshots into square boxes of size $15\times15$ and compute both nematic order parameter and packing fraction within the cutouts.
Aggregated over space, time, initial conditions, and both circular and channel domains, we show point clouds of the observed order--packing fraction pairs in \cref{pan:pointclouds} for selected parameters.

For strongly oblate and prolate rods we observe a wide range of order parameters and the most efficient packing at perfect order.
Note that due to their construction as described in the model section, the corner radii decrease for smaller $\ad <1$.
Due to this and to compression effects, the packing fraction, or, more accurately, the area density may slightly exceed 1 for perfect order in strongly oblate rods.
Rods with intermediate aspect $\ad$, on the other hand, only generate partial order, albeit tunable with the shear flow.
For these, the largest packing fraction is observed for vanishing or intermediate order parameter.
Ellipses show a very similar picture across all $\ad$.
The range of observable order parameter values depends on the domain geometry, more specifically, the imposed shear flow, and the most packing-efficient order parameter is observed at $\ad$-dependent partial order.

These data are further condensed into histograms drawn as order--aspect heatmaps for elliptical particles in \cref{pan:packing_heatmap_ellipse} and for rods in \cref{pan:packing_heatmap_rod}.
White regions indicate that no data were observed.
We find a remarkable connection between these histograms and the average order parameter observed in channels from \cref{fig:fig2} (here drawn in red). For prolate particles, the order parameter is always near or greater than the value with most efficient packing. The same is true for oblate elliptical particles, but not for oblate rods.
This shows that the packing fraction function can indeed be a large puzzle piece in understanding the emergent order parameters but by itself, it does not capture the kinematic inhibition of global average order in oblate rods due to the lack of rearrangements. This is what we will address below.

\subsection{Order-dependent viscous relaxation}

Strongly interdependent with the configurations of particles and their shapes is the dynamics that arranges them.
In \cref{pan:stats_excess_vel} we showed that particle shapes can impact the expansion flow at the mesoscopic scale in some cases leading to significant deviations from the expected mean-field flow.

However, this is not the case for all shapes and the observed dynamics at mesoscopic scales must be driven by differences at the microscopic level which we will study here.
Each particle grows along its orientation direction.
Naively, if all particles were firmly attached to their neighboring particles, each patch of particles should expand according to its local order parameter - isotropically when fully disordered and fully directed when maximally ordered.
However, driven by growth, particles continuously rearrange in their local environment leading to deviations from this naive assumption and here we study these local dynamics to characterize them as coarse-grained viscous effects.

Based on simulation snapshots separated by $\Delta t = 0.1$ we compute the finite difference velocity of all particles that exist in both snapshots.
Then we compute the particle velocities $\bm{v}_{\mathrm{rel}}$ for each particle in the co-moving and co-rotating frame of its 20 nearest neighbors.
It has zero-mean and its average variation, decomposed along $v_{\parallel} = \bm{v}_{\mathrm{rel}}\cdot\bm{n}$ and orthogonal $v_{\perp} = \bm{v}_{\mathrm{rel}}-\bm{n}v_{\parallel}$ to the local average orientation $\bm{n}$, is shown in \cref{pan:local_vel_stds}.
An illustration of the relative particle motion is shown in \cref{pan:local_vel_rodsketch} and \cref{pan:local_vel_ellipsesketch} where the fundamental difference in directional statistics already becomes apparent. 
Quantitative results are shown for both rod-shaped and elliptical particles with varied division aspect $\ad$. For each combination, we show a dashed line computed from samples with smallest scalar order $\xi_{\mathrm{local}}$ (lowest $10\%$ of samples), and an approximation from the top $1\%$ samples is shown as a solid line.
The former is mostly for reference and the latter is most relevant, as it has the potential for generating the most anisotropic expansion.
The corresponding local order average of the top $1\%$ is shown in \cref{pan:local_vel_order} which reveals large values for both strongly oblate and prolate division aspect $\ad$.
Near the transition from oblate to prolate shapes there is a region where only partial order is observed in both models though narrower and slightly shifted right for the rod-shape.

Two derived quantities that shed insight into the effective coarse-grained dynamics are shown in \cref{pan:local_vel_ratio} and \cref{pan:local_vel_min}.
The ratio of velocity variations shown in \cref{pan:local_vel_ratio} for elliptical particles follows the ensemble aspect ratio $\ea$ very closely.
This is an important insight.
In the case of strong local ordering, as observed in \cref{pan:local_vel_order}, the ensemble aspect $\ea$ matches the anisotropic granularity of the particle patch.
As a consequence, taking the anisotropic granularity into account we find the remaining local velocity variations becoming fully isotropic and independent of orientation direction for both oblate and prolate elliptical particles.

The same is also approximately true for rod-like particles with $1 < \ad < 4$ but strongly anisotropic behavior remains outside of that regime.
We propose to interpret this observation in the following way:
Viscous relaxation of strains boils down to neighbor changing rearrangements of particles at the microscopic level such that the ensemble of fixed-shape particles collectively occupies space of a different shape i.e. turning vertical stacking into horizontal stacking.
For isotropic particles this requires approximately equal amounts of displacement along the direction of compression and orthogonal to it.
For our anisotropic but ordered particles we can therefore expect an appropriately anisotropic contribution on top of additional directed growth rate effects.
To estimate the rate of local viscous relaxation, we therefore compute
\begin{align}
    \text{viscous flux} = \min\left(\sqrt{\langle v_{\parallel}^2\rangle\ea^{-1}}, \sqrt{\langle v_{\perp}^2\rangle \ea}\right)\label{eq:viscflux}
\end{align}
explicitly using that viscous relaxation requires equal contributions along both axes after rescaling for particle aspect ratio.
This is shown in \cref{pan:local_vel_min}.
In this interpretation, we find the rate of viscous relaxation to be approximately independent of division aspect $\ad$ for elliptical particles, albeit with a pronounced step at the transition between fully oblate and prolate shapes.

Rod-shaped particles on the other hand show significant suppression of the viscous relaxation in the highly ordering regime for both oblate and prolate aspect ratios.
The anisotropic strain introduced by directed growth must therefore be isotropized over greater length scales, giving rise to the microdomain dynamics and increased deviations from the mean-field flow seen in \cref{pan:stats_excess_vel}.
The combination of order-dependent packing fractions controlling the emergence of order and similarly order-dependent viscous relaxation describing the response to directed growth and shear finally paves the way for a mechanistic understanding of these rich shape-dependent dynamics.
While elliptical particles find ordered states due to advantageous packing fractions, viscous relaxation can compensate growth-induced strain continuously. For rod-shaped particles on the other hand, the ordered configurations will, as measured, inhibit viscous relaxation locally leading to directed expansion of forming microdomains and thus the characteristic microdomain dynamics.

\begin{figure}
    \centering
    \includegraphics[width=\linewidth]{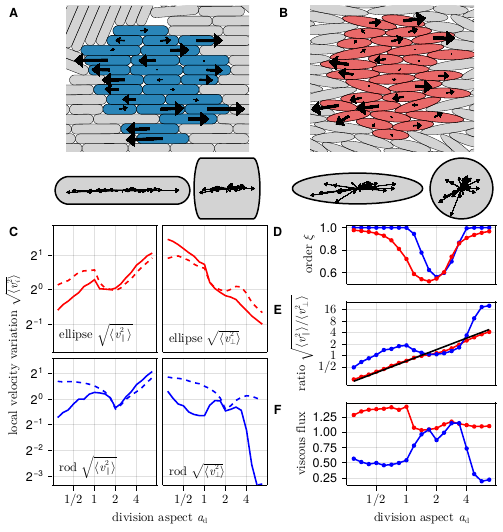}
    \caption{\label{fig:local_vel_variance}%
    \panel{pan:local_vel_rodsketch} Rod velocities as arrows in co-moving frame with additional visualization of velocities centered in single particle and with rescaled coordinates.
    \panel{pan:local_vel_ellipsesketch} Analogous to \ref{pan:local_vel_rodsketch} for elliptical particles.
    \panel{pan:local_vel_stds} Local velocity variations co-aligned with nematic order and relative to co-moving frame for elliptical and rod-shaped particles as labelled.
    Averages over samples with lowest $10\%$ order (dashed) and highest $1\%$ order (solid).
    \panel{pan:local_vel_order} Average of locally observed order (top $1\%$)
    \panel{pan:local_vel_ratio} Ratio of velocity variations (top $1\%$) with ensemble aspect $\ea$ added in black.
    \panel{pan:local_vel_min} Viscous flux approximation according to \cref{eq:viscflux}.
    }
\end{figure}

\section{Discussion and Conclusion}

We presented agent-based simulations of proliferating, sterically interacting anisotropic particles (rounded rectangles and ellipses) in two classes of domains and systematically varied their division aspect to span regimes where flow-induced alignment and division geometry either cooperate or compete.
The main result is that changes in particle shape qualitatively change collective behavior: (i) rounded ellipses follow classical shape-based flow-alignment even in the oblate regime leading to growth being predominantly directed perpendicular to imposed channel flow, (ii) flat-sided oblate rods reproduce the familiar microdomain dynamics known from prolate rods and associated stress anisotropies, and (iii) ellipses generate large-scale orientational correlations without forming stable microdomains, leading to efficient packing at imperfect nematic order.

Mechanistically, these outcomes follow from a competition between two alignment cues.
Growth and binary division provide a natural source of alignment of particles along their growth axis, while shear and crowding favor (shape) alignment with the mean shear flow.
Tip curvature and side flatness determine the ease with which neighbors can slide and rearrange; flat sides stabilize local parallel alignment and microdomains, whereas smoothly rounded shapes enable frequent neighbor exchange and relaxation of anisotropic strains.
Quantitatively, the signed order parameter flips sign near a division aspect $\ad\approx4/3$, stress anisotropy in channels depends non-monotonically on $\ad$, and orientation correlation lengths show a strong dependence on shape, reflecting distinct relaxation pathways in ellipses versus rods.

These particle-scale observations motivate a compact coarse-grained description in which an order-dependent packing fraction and an effective, order-dependent particle aspect ratio act as principal state variables that couple to growth-driven density increase and anisotropic stress. Condensing microscopic shape and rearrangement rules into these two fields yields minimal continuum ingredients that capture the reversal of flow-alignment, the emergence and breakup of microdomains, and the distinct relaxation dynamics we observe.

Practically, the order-dependent packing fraction $φ(\xi)$ could be inserted into continuum descriptions as a constitutive input in order-stress couplings to predict when and at which scale ordered domains form or decay, thereby providing a directly measurable bridge from particle-resolved simulations to predictive coarse-grained theory.

Beyond the immediate modeling advances, our results emphasize a simple physical point: geometric details that affect local rearrangement kinetics can control large-scale organization in proliferating active matter.
This suggests experimental tests in microbial or cellular systems where cell shape or division orientation can be tuned and motivates extensions to three dimensions, to systems with turnover (death or extrusion), and to models that include hydrodynamic or biochemical couplings.

In summary, this study separates and then recombines two dominant alignment mechanisms: alignment driven by growth and division, and alignment induced by shear flow. This approach identifies a small set of geometric and kinematic parameters that control collective order in proliferating assemblies. The resulting predictions are compact and robust, and provide a roadmap for unifying particle-resolved simulations with continuum theories of growing, anisotropic active matter.

\section*{Data Accessibility}
Simulation codes used for this study were built on top of the open-source framework InPartS.jl~\cite{hupeInPartSInteractingParticle2022a} with analyses conducted using the julia programming language~\cite{bezansonJuliaFreshApproach2017} and visualizations using Makie.jl~\cite{DanischKrumbiegel2021}.
The implementation of the agent-based models will be made available alongside the final publication.\\

\acknowledgments
We thank Ramin Golestanian and the Department of Living Matter Physics at MPI-DS for their support. We acknowledge support from the Max Planck Society as well as the Max Planck School Matter to Life, which is jointly funded by the Federal Ministry of Education and Research (BMBF) of Germany and the Max Planck Society.

\newpage
\appendix
\section{Repulsion forces}
\label{app:repulsionforces}
Elastic bodies deform when they are pushed together.
In our modeling scheme we require that all effective deformations are small compared to the particle sizes.
This allows us to describe the deformations implicitly and instead consider small overlaps between particles that cause repulsion forces.

We first introduced rounded rectangular particles in Ref.~\cite{isenseeSensitiveParticleShape2025}, although with a different parameter regime in mind.
Nonetheless, the implementation remains identical.
To compute the interaction force between two rounded rectangles, we find the closest points on the rectangular frame construction.
Based on the closest points, we apply the repulsion force
\begin{equation}
  \label{eq:model_rod_external_steric}
 \bm{F}_{ij}(\bm{\delta}_{ij}) = \frac{Y}{2}\sqrt\frac{R_c}{2}\,\qty(2R_c - \norm{\bm{\delta}_{ij}})^{3/2} \,\frac{\bm{\delta}_{ij}}{\norm{\bm{\delta}_{ij}}}
\end{equation}
with Hertzian scaling, prefactor $Y$, (corner) radius $R_c$, and connecting vector $\bm{δ}$ between the closest points on either particle.

For elliptical particles, the analogy of closest points on finite line segments cannot be applied.
Instead, we find the intersection points of touching ellipses which can be computed as the roots of a quartic polynomial.
This is done using an optimized custom implementation that also conducts plausibility checks on the solution to combat the well-known catastrophic cancellation problems occurring for ellipse pairs that overlap by amounts on the order of the floating-point precision.
Given a pair of intersection points, we apply a repulsion force normal to their connecting vector and proportional to their distance.
\end{document}